\documentclass[12pt,letterpaper]{article}
\usepackage[latin1] {inputenc} 
\usepackage{amsmath} \usepackage{amsfonts} \usepackage{amssymb} 
\begin{document}

\title{DSR and Canonical Transformations: A Comment on a ``A Lagrangian for DSR particle and the role of noncommutativity''}
\author{J. Antonio Garc\'\i a\footnote{email:
garcia@nucleares.unam.mx} \\
\em Instituto de Ciencias Nucleares, \\
\em Universidad Nacional Aut\'onoma de M\'exico\\
\em Apartado Postal 70-543, M\'exico D.F., M\'exico}
\maketitle
\abstract{The aim of this comment is to call to the attention of DSR readers a basic
fact.  The introduction of
noncommutative structures in problems like the one addressed in [1] is not
necessary for the understanding of DSR physics. It can be described just as
the relativistic free particle problem in a different parametrization.}

\section*{}

An important task of mathematical physics is to prove that a given mathematical 
description of a physical theory is under some mild conditions ``unique".
This is a difficult problem that can be tackled in principle with tools like consistent
deformation, Lie algebra stability analysis and cohomology. A fundamental requirement
for a good description of a  physical theory is its invariance under field redefinitions. 
For example, one of the basic aims of the 
variational formulation in classical mechanics is precisely its covariance in 
configuration space.

It is important not to confuse the physical setup with its description. For instance we can start with the free
relativistic particle and use any appropriate set  of variables to describe its law of motion that, 
of course, is ever the same; or  we can start from the harmonic oscillator and use the Hamilton-
Jacobi description but this does not imply that the law of motion is $\dot Q=0, \dot P=0$ 
in phase space.

Two different descriptions of the same physics are such 
that they can be related
through  an invertible transformation that can be a canonical  transformation or a general ``field redefinition''. They represent the same physical content in a different parametrization. 

Based on these observations we would like to comment on the result obtained in \cite{Ghosh}
where the author propose a ``new" Lagrangian in configuration space for the 
free relativistic particle
in Double Special Relativity (DSR).

We will show that this new Lagrangian can be related to the Lagrangian
of the free relativistic particle through a very simple canonical transformation.
In fact, we will show that a lot of models (in the classical framework) including
some of the so called $\kappa-$deformed\footnote{The class of systems that we are addressing here are relativistic free particles with modified dispersion relations and with a trivial coproduct. In order to be precise these must be considered as trivial $\kappa$-deformations. Thanks to the referee for his comment about this point.}  ones, can be related also with the Lagrangian of 
the free particle of the standard special relativity.

For simplicity, start from the Lagrangian of the free relativistic particle of mass $m$
\begin{equation}\label{L}
L=\dot X^\mu P_\mu- \frac{e}{2}(P^2-m^2).
\end{equation}
Using the equations of motion for $P$, we can eliminate the momenta from the given Lagrangian
\begin{equation}\label{Le}
L=\frac{\dot X^2}{2e}+\frac{e}{2}m^2,
\end{equation}
and by the same token we can eliminate also the Lagrange multiplier $e$
to get the Lagrangian in configuration space $X$:
\begin{equation}\label{Ldotx}
L=m\sqrt{\dot X^2}.
\end{equation}
Now implement in (\ref{L}) the redefinition of momenta $P_\mu\to f_\mu(p)$  where $f_\mu$ is an arbitrary function with a well-defined inverse. This redefinition can always be completed to a canonical transformation \cite{Gold} whose generator is
$$F_3(X,p)=-f_\mu X^\mu,$$
so the transformation rule for the $X^\mu$ is
$$X^\mu= (F^{-1})^\mu_\nu x^\nu,$$
where 
$$F^\mu_\nu=\frac{\partial f_\nu}{\partial p_\mu}.$$
The new DSR Lagrangian is
\begin{equation}\label{LDSR}
L_{DSR}=\dot x^\mu  p_\mu - e(f^2-m^2) .
\end{equation}
As the transformation is canonical, the Poisson algebra of the Lorentz group is unchanged but the generators changes to
$$J^{\mu\nu}=X^\mu P^\nu- X^\nu P^\mu\to J^{\mu\nu}_{DSR}= (F^{-1})^\mu_\rho x^\rho f^\nu-(F^{-1})^\nu_\rho x^\rho f^\mu.$$
To obtain from here the MS generators \cite{MS} and the new Lagrangian claimed in \cite{Ghosh}, just use
$$f_\mu=\frac{p_\mu}{1-\ell p_0},$$
into (\ref{LDSR}) and the same procedure outlined to obtain (\ref{Ldotx}) from (\ref{L}). An example of a  $\kappa-$deformed model \cite{KG} can be obtained from
$$f_0=\kappa\sqrt{\cosh (p_0/\kappa)}, \quad f_i=\exp{(p_0/\kappa)}\frac{p_i}{2}.$$

Notice that we do not need any noncommutative ansatz and/or exotic Dirac brackets to fix in an {\em ad hoc} way the resulting Lorentz algebra. It is clear from our analysis that you can also play other games with more general field redefinitions. The point is that, when you chose one field redefinition, you must be consistent by applying it to the Lagrangian and also to the Lorentz generators. 
In particular, the noncommutative structure constructed in eq. (10) of  \cite{Ghosh} using the Dirac
algorithm for constrained dynamical systems, by a gauge fixing procedure
used to reproduce in {\em ad hoc} way through the Dirac Bracket the
noncommutative structure that he wants, is a fake noncommutativity. 
All the procedure to obtain this noncommutative
symplectic structure
can be mapped to the relativistic free particle problem using the same canonical
transformation.
The apparent contradiction come from the fact that a canonical transformation
preserves the symplectic structure of the original phase space, but clearly does
not preserve the Dirac bracket.

As an aside, this noncommutative structure  is not consistent
with the new Lagrangian (see eq. (1) in \cite{Ghosh}) in configuration space \cite{CG}.

Of course, we are not claiming that this result invalidates all of the analyses of DSR
and $\kappa-$deformed physics. This problem must be tackled from the physical
setup and not from the description. What we are claiming is that there exist a very
easy framework to describe the free relativistic particle in DSR and some $\kappa-$deformed scenarios, including the problems associated with the physical 
interpretation
that can also be mapped  with the help of the canonical transformation. Our approach is modest
even though not trivial. We are just calling
attention to one basic issue: the analysis of the physics behind DSR from
the perspective of the {\em classical physics} could be an incompletely
defined problem (without additional information, e.g. about the
noncommutativity of the geometry), because it can be tackled with standard and
very well-known techniques.

{\bf Update}: While this comment was in the editorial process at PRD, an interesting work about the relation of DSR with canonical transformations  appeared \cite{galan}. It has some intersection with the ideas presented here and also has a representative set of references that reflect the current state of the problem addressed here.   
A paper with the same contents as the one commented on here was also published in \cite{Ghosh2} where an incorrect deduction of the Lagrangian in configuration space (eq. (1) of \cite{Ghosh}) was presented.

The author acknowledge enlighten discussions with David Vergara and support from  grants CONACyT 32431-E and DGAPA IN104503.

\end{document}